\documentclass[10pt,conference]{IEEEtran}
\usepackage{amsmath,amssymb,bm}
\usepackage[pdftex]{graphicx}

\newcommand{\Const}{\mathrm{Const}}

\newcommand{\Seg}{\mathrm{Seg}}

\newcommand{\mini}{\mathrm{minimize \ }}
\newcommand{\st}{\mathrm{\quad s.t. \ }}
\newcommand{\Fc}{\mathcal{F}}
\newcommand{\Hc}{\mathcal{H}}
\newcommand{\Pc}{\mathcal{P}}
\newcommand{\Kc}{\mathcal{K}}
\newcommand{\Vc}{\mathcal{V}}
\newcommand{\Rc}{\mathbb{R}}
\newcommand{\df}{d_{\mathrm{frac}}}
\newcommand{\dfa}{d^{\mathrm{after}}_{\mathrm{frac}}}
\newcommand{\Supp}{\mathrm{Supp}}
\newcommand{\U}{\mathcal{U}}

\newcommand{\Ffinact}{\mathbf{F}_{\mathrm{inact}}}
\newcommand{\Ffact}{\mathbf{F}_{\mathrm{act}}}
\newcommand{\rlx}{^{(\mathrm{relax})}}

\newtheorem{defi}{Definition}
\newtheorem{thim}{Theorem}
\newtheorem{prf}{Proof}
\newtheorem{lem}{Lemma}
\newcommand{\qed}{\fbox{}}
\begin{document}

% paper title
%\title{Improving Fractional Distance of Binary Linear Codes Assign Redundant Parity Check Matrix}
\title{A Cutting Plane Method based on Redundant Rows for Improving Fractional Distance}

\author{\authorblockN{Makoto Miwa\authorrefmark{1},
Tadashi Wadayama\authorrefmark{1} and
Ichi Takumi\authorrefmark{1}}
\authorblockA{\authorrefmark{1}Graduate School of Engineering\\
Nagoya Institute of Technology,\\
Gokiso-cho, Showa-ku, Nagoya, 466-8555 Japan\\
Email: mkt-m@ics.nitech.ac.jp, wadayama@nitech.ac.jp, takumi@ics.nitech.ac.jp}}

\maketitle
\begin{abstract}
In this paper, an idea of the cutting plane method is employed 
to  improve the fractional distance of a given binary parity check matrix.
The fractional distance is the minimum weight (with respect to $\ell_1$-distance) of vertices
of the fundamental polytope. 
The cutting polytope is defined based on redundant rows of the parity check matrix 
and it plays a key role to eliminate unnecessary fractional vertices in the fundamental polytope.
We propose a greedy algorithm and its efficient implementation for improving the fractional distance 
based on the cutting plane method.
\end{abstract}

\section{Introduction}

Linear programming (LP) decoding proposed by Feldman \cite{Feldman}
is one of the promising decoding algorithms for low density-parity check (LDPC) codes.
The invention of LP decoding opened
a new research field of decoding algorithms for binary linear codes. 
Recently,  a number of studies on LP decoding have been made such as 
\cite{Mohammad}\cite{Kelley}\cite{Ralf}.

LP decoding has several virtues that belief propagation (BP) decoding does not possess. 
One of the advantages of LP decoding is that the behavior of the LP decoder can be clearly understood
from a viewpoint of optimization.
The decoding process of an LP decoder is just a minimization process of a linear function 
subject to linear inequalities corresponding to the parity check conditions. The feasible set of 
this linear programming problem is called  the {\em fundamental polytope} \cite{Feldman}.
The fundamental polytope is a relaxed polytope that includes the convex hull of all the codewords of
a binary linear code. Decoding performance of LP decoding is thus closely related to 
geometrical properties of the fundamental polytope.

Another advantage of LP decoding is that its decoding performance can be 
improved by including additional constraints in the original LP problem.
The additional constraints tighten the relaxation and they lead to improved decoding performance.
Of course, additional constraints increase the decoding complexity of LP decoding but 
we  can obtain  flexibility to choose a trade-off  between performance and complexity.
The goal of the paper is to design an efficient method to find
additional constraints that improve this trade-off for a given parity check matrix.

In this paper, we will propose a new greedy type algorithm to enhance the fractional distance
based on the cutting polytope based on the redundant rows 
that eliminates unnecessary fractional vertices in the fundamental polytope.
The additional constraints generated by the proposed method  is based on
redundant rows of a parity check matrix. In other words, the proposed method can be considered
as a method to generate redundant parity check matrices from a given original parity check matrix.
Therefore, the present work has close relationship to the works on 
elimination of stopping sets (SS) by using redundant parity check matrices and on 
stopping redundancy such as \cite{Schwartz} \cite{ghaffar} \cite{Hollmann}.

The cutting plane method is a well established technique for solving 
an integer linear programming (ILP) problem based on LP \cite{Papa}.
The basic idea 
of the cutting plane method is simple. In the first phase, an ILP problem is relaxed to 
an LP problem and then it is solved by an LP solver. If we get a fractional solution 
(i.e., a vector with elements of fractional number), 
a cutting plane
(i.e., an additional  linear constraint) matched to the fractional solution is added to the LP problem in the second phase.
The cutting plane is actually  a half space which excludes the fractional solution but it includes 
all the ILP solutions. In the third phase,  the extended LP problem is solved and the above process 
continues until  we obtain an integral solution.

The {\em fractional distance} of a binary linear code is  the $\ell_1$-weight 
of the minimum weight vertex of the fundamental polytope. The fractional distance  is known to be 
an  appropriate geometrical property which indicates the decoding performance of 
LP decoding for the binary symmetric channel (BSC). 
It is proved that the LP decoder can successfully correct bit flip errors if the number of errors is less than 
half of the fractional distance \cite{Feldman}. 
In contrast to the minimum distance of a binary linear code, we can evaluate the fractional distance
efficiently with an LP solver.  Efficient evaluation of the fractional distance is especially important 
in the proposed method.

The idea for improving LP decoding performance using redundant rows of 
a given parity check matrix was discussed in  \cite{Mohammad} \cite{Feldman2}.
In their methods, the redundant rows  are efficiently found based on a short cycle of a Tanner graph of
the parity check matrix.
Their results indicate that addition of redundant rows is a promising technique to 
improve LP decoding performance and they suggest that further studies on this subject are meaningful 
to pursue more systematic ways to find appropriate redundant 
rows that achieves better trade-offs between decoding performance and complexity.

\section{Preliminaries}
In this section, notations and definitions required throughout the paper 
are introduced.
\subsection{Notations and definitions}
Let $H$ be a binary  $m \times n$ matrix where $n > m \ge 1$.
The binary linear code defined by $H$,  $C(H)$, is defined by
\begin{equation}
	C(H) \equiv \left\{ \bm{x} \in \Bbb F_2^n : \bm{x} H^T = 0^m \right\},
\end{equation}
where $\Bbb F_2$ is the Galois field with two elements $\{0,1\}$ and 
$0^m$ is the zero vector of length $m$.
In the present paper, 
the bold face letter, like $\bm{x}$, denotes a row vector.
The elements of a vector is expressed by corresponding normal face letter with subscript;
e.g., $\bm{x} = (x_1,x_2,\ldots, x_n)$.
The following definition of the fundamental polytope is due to Feldman \cite{Feldman}.
\begin{defi}{(Fundamental polytope)}
Assume that $\bm{t}$ is a binary row vector of length $n$.
Let 
\begin{equation}
	X(\bm{t}) \equiv \{S \subseteq \Supp(\bm{t}) : |S| \ \mathrm{is \ odd} \},
\end{equation}
where $\Supp(\bm{t})$ denotes the support set of the vector $\bm{t}$ defined by
\begin{equation}
	\Supp(\bm{t}) \equiv \{i \in \{1,\ldots, n\}:  t_i \ne 0  \}.
\end{equation}
The single parity polytope of $\bm{t}$ is the polytope defined by 
\begin{eqnarray}
	\U(\bm{t}) \hspace{-3mm}& \equiv &\hspace{-3mm} \{ \bm{f} \in [0,1]^n :  \forall S \in X(\bm{t}), \nonumber \\
	& & \hspace{-8mm} \sum_{j \in S} f_j + \sum_{j \in (\Supp(\bm{t}) \setminus S)} (1 - f_j) \le |\Supp(\bm{t})| - 1 \},
\end{eqnarray}
where $[a, b] \equiv \{ v  \in \Rc : a \le v \le b \}$. The symbol $\Rc$ denotes the set of real numbers.
For a given binary $m \times n$ parity check matrix $H$, 
the fundamental polytope $\Pc(H)$ is defined by
\begin{eqnarray}
	\Pc(H) \equiv \bigcap_{i=1}^m \U(\bm{h}_i), \label{equ:PcH}
\end{eqnarray}
where $\bm{h}_i = (h_{i1}, h_{i2}, \ldots, h_{in})$ is the $i$-th row vector of $H$.
\hfill\qed
\end{defi}

The convex hull of $C(H)$ is the intersection of all convex sets 
including $C(H)$\footnote{In this context, we implicitly assume that $C(H)$ is embedded in $\Rc^n$.}.
It is known that the fundamental polytope $\Pc(H)$ contains the convex hull of $C(H)$ as a
subset and that $\Pc(H) \cap \{ 0, 1 \}^n = C(H)$ holds.

Let $M$ denote the number of the linear constraints (inequalities) that forms $\Pc(H)$.
Assume that these linear constraints are numbered from $1$ to $M$
and that the $k$-th linear constraint has the form:
\begin{equation}
	\alpha_{k1} f_1 + \cdots + \alpha_{kn} f_n \leq \beta_k,
\end{equation}
for $k \in \{ 1, \ldots, M \}$.
We call the $k$-th constraint  $\Const_k$.
The hyper plane corresponding to  $\Const_k$ is given by
\begin{equation}
	\Fc_k \equiv \{ \bm{f} \in \Rc^n : \alpha_{k1} f_1 + \cdots + \alpha_{kn} f_n = \beta_k \},
\end{equation}
and the half space satisfying $\Const_k$ is defined by
\begin{eqnarray}
	\Hc_k \equiv \{ \bm{f}\in \Rc^n : \alpha_{k1} f_1 + \cdots + \alpha_{kn} f_n \leq \beta_k \}.
\end{eqnarray}
The fundamental polytope $\Pc(H)$ is thus the intersection of the half spaces such that
\begin{equation}
	\Pc(H) = \Hc_1 \cap \Hc_2 \cap \cdots \cap \Hc_M.
\end{equation}
Let $\Ffact(H)$ be the set of indices of the active constraints defined by
\begin{equation}
	\Ffact(H) \equiv \{ k \in \{ 1, \ldots, M \} : 0^n \in \Fc_k \}.
\end{equation}
The active constraints are the linear constraints 
whose hyper plane contains the origin $0^n$.
In a similar manner, we define 
$\Ffinact(H)$, which is the set of indices of the inactive constraints, by
\begin{equation}
	\Ffinact(H) \equiv \{ k \in \{ 1, \ldots, M \} : 0^n \notin \Fc_k \}.
\end{equation}
The fundamental cone $\Kc(H)$ is  the cone defined by the active constraints:
\begin{eqnarray}
	\Kc(H) \equiv \{ \bm{f} \in [0,1]^n : \forall k \in \Ffact(H), \ \bm{f} \in \Hc_k \}. \label{equ:cone}
\end{eqnarray}

\section{Cutting plane method}
In this section, the main idea of the cutting plane method based on redundant rows 
will be introduced and then an application to Hamming code is shown as an example.
\subsection{Fractional distance}
The fractional distance $\df(H)$ is the $\ell_1$-distance 
between a codeword and the nearest vertex of $\Pc(H)$ \cite{Feldman}. 
\begin{defi}{(Fractional distance)}
Let $\Vc(X)$ be the set of all vertices of a polytope $X$.
For a given binary $m \times n$ parity check matrix $H$, 
the fractional distance of $H$ is defined by
\begin{equation}
	\df(H) \equiv \min_{
		\begin{subarray}{c}
			\bm{x} \in C(H) \\
			\bm{f} \in \Vc(\Pc(H)) \\
			\bm{x} \ne \bm{f}
		\end{subarray}
		} \sum_{i=1}^n |x_i - f_i|. \vspace{-2mm} 
\end{equation}
\hfill\qed
\end{defi}

The importance of the fractional distance is stated in the following lemma due to 
Feldman \cite{Feldman}.
\begin{lem}
Assume that the channel is BSC.
Let $e$ be the number of the bit flip errors.
If 
\begin{equation}
	\left\lceil \frac{\df(H)}{2} - 1 \right\rceil \ge e,
\end{equation}
holds, then 
all the bit flip errors can be corrected by the LP decoder. \\
(Proof) The proof is given in \cite{Feldman}.
\hfill \qed
\end{lem}
Based on the  geometrical uniformity of the fundamental polytope (called {\em C-symmetry} in \cite{Feldman}), 
it has been proved that 
$\df(H)$ is the $\ell_1$-weight of the minimum weight vertex of $\Pc(H)$ except for the origin, 
which is expressed by
\begin{equation}
	\df(H) = \min_{
			\bm{f} \in \Vc(\Pc(H)) \setminus 0^n
		} \sum_{i=1}^n f_i. \label{equ:def_frac}
\end{equation}
Let $\Gamma(H)$ be a set of the minimum weight vertices of $\Pc(H)$:
\begin{eqnarray}
	\Gamma(H) \equiv \left\{ \bm{p} \in \Vc(\Pc(H)) : \sum_{i=1}^n p_i = \df(H) \right\},
\end{eqnarray}
and let $d_{\min}$ be the minimum distance of $C(H)$. Since any codeword 
of $C(H)$ is a vertex of $\Pc(H)$, it is obvious that the inequality 
$\df(H) \le d_{\min}$ holds for any $H$. 
The fractional distance $\df(H)$  depends on the representation of a given binary linear code
(i.e., the parity check matrix) and 
there are a number of parity check matrices that define  the same  binary linear code.
This means that  the parity check matrices of a binary linear code can be ranked 
in terms of its fractional distance. It is hoped to find a better parity check which achieves larger 
factional distance for a given binary linear code because such a parity check matrix 
may improve the performance/complexity trade-off of LP decoding for the target code.

\subsection{Cutting polytope}
In this subsection, we will define the cutting polytope based on the redundant rows.
The next definition gives the definition of the {\em redundant row}.
\begin{defi}{(Redundant row)}
Let $H$ be a binary $m \times n$ parity check matrix of the 
target code $C$. A redundant row $\bm{h}$ is a linear combination of the row vectors of $H$ such that
\begin{equation}
	\bm{h}  = a_1 \bm{h}_1 + a_2 \bm{h}_2 +  \cdots + a_m \bm{h}_m,
\end{equation}
where $a_i \in \Bbb F_2 (i \in \{1,\ldots, m\})$.
\hfill\qed
\end{defi}
The single parity polytope of a redundant row $\bm{h}$ 
includes all the codewords of $C(H)$ because any codeword $\bm{x} \in C(H)$ satisfies
$\bm{x} \bm{h}^T = 0$ \cite{Schwartz}.

The next lemma asserts a cutting property of a single parity polytope satisfying 
a certain condition.
\begin{lem}\label{lem:exclude}
Let $H$ be a binary $m \times n$ parity check matrix.
Assume that  a point $\bm{p} \equiv ( p_1, \ldots, p_n) \in \Rc^n$ and
$\bm{t} \in \Bbb F_2^n$ are given.
If
\begin{equation} 
\exists j \in \Supp(\bm{t}), \quad p_j > \sum_{l \in \Supp(\bm{t}) \setminus \{j \}} p_l, \label{equ:h_add}
\end{equation}
holds, then $\bm{p} \notin \U(\bm{t})$ holds. \\
(Proof) Let $j^*$ be the index satisfying 
\begin{equation}
p_{j^*} > \sum_{l \in \Supp(\bm{t}) \setminus \{j^* \}} p_l.
\end{equation}
The above inequality is equivalent to the following inequality:
\begin{equation}
\sum_{j \in \{ j^* \}} p_j + \sum_{l \in \Supp(\bm{t}) \setminus \{j^*\}}(1- p_l) > |\Supp(\bm{t})| - 1.
\end{equation}
From the definition of $\U(\bm{t})$, it is evident that $\bm{p} \notin \U(\bm{t})$.
\hfill\qed
\end{lem}

The cutting polytope defined below is used to cut a fractional vertex 
in the proposed method for improving the fractional distance.
\begin{defi}{(Cutting polytope)}
Assume that $\bm{p} = (p_1,p_2,\ldots, p_n) \in \Pc(H)$. 
If a redundant row $\bm{h}$ with the form:
\begin{equation}
\bm{h}  = a_1 \bm{h}_1 + a_2 \bm{h}_2 +  \cdots + a_m \bm{h}_m
\end{equation}
satisfies 
\begin{equation}
	p_j > \sum_{l \in (\Supp(\bm{h}) \setminus \{j \})} p_l, \label{equ:cnd}
\end{equation}
where $j = \arg \max_{i \in \Supp(\bm{h})} p_i$,
then $\U(\bm{h})$ is called a cutting polytope of $\bm{p}$.
\hfill\qed
\end{defi}

\begin{figure}[ht]
	\centering
	\includegraphics[scale=0.45]{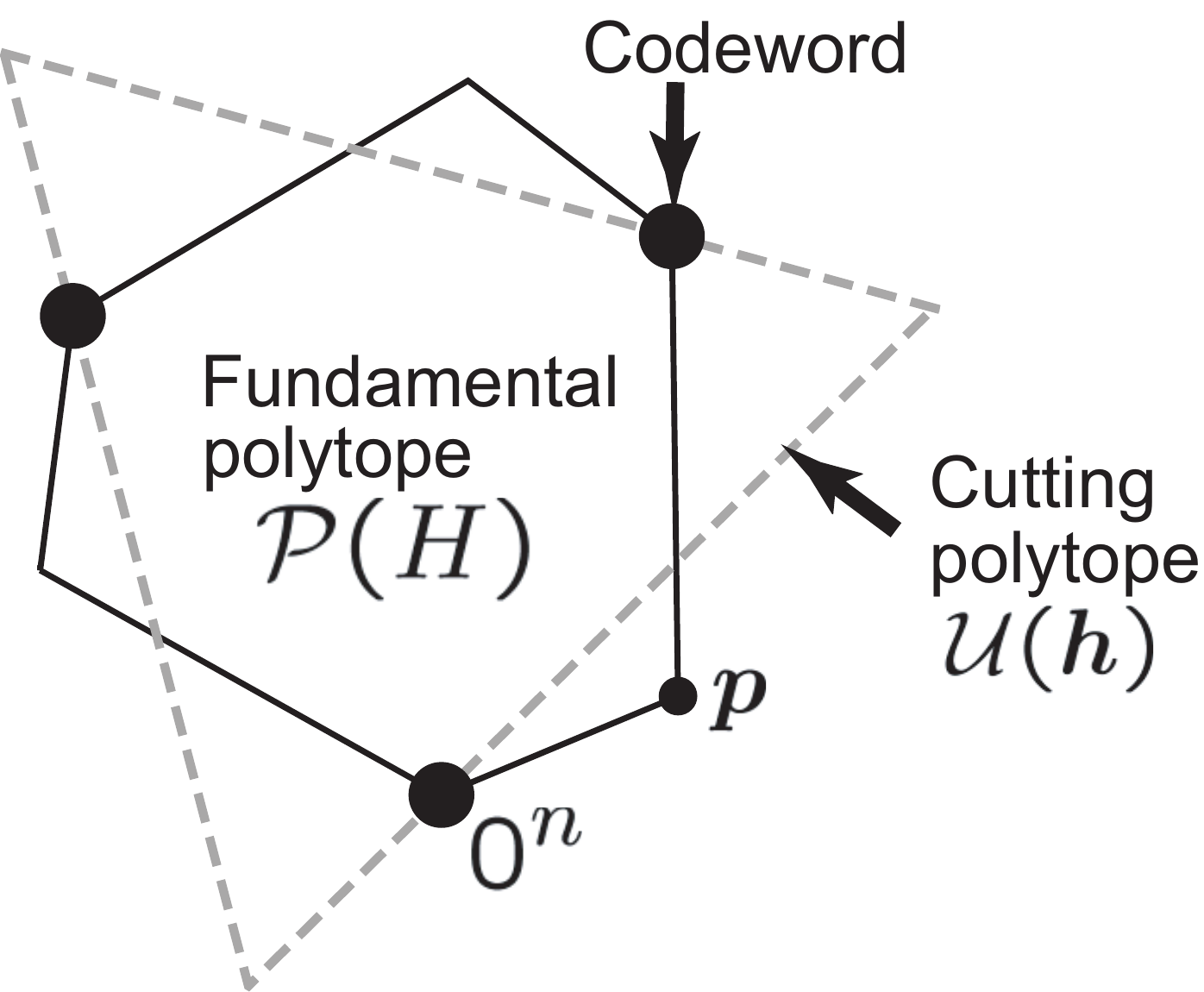}
	\caption{The idea of the cutting polytope}
	\label{fig:cutting}
\end{figure}

The next theorem introduce a tighter relaxation of the convex hull of $C(H)$ which may improve
the fractional distance.
\begin{thim}
Let $\bm{p} = (p_1,p_2,\ldots, p_n) \in \Pc(H)$ and 
$\U(\bm{h})$ be a cutting polytope of $\bm{p}$.
The following relations hold:
\begin{equation}
C(H) \subset \Pc(H) \cap \U(\bm{h})
\end{equation}
and 
\begin{equation}
\bm{p} \notin \Pc(H) \cap \U(\bm{h}).
\end{equation}
(Proof) The claim of the theorem is directly derived from $C(H) \subset \U(\bm{t})$ and Lemma \ref{lem:exclude}.
\hfill\qed
\end{thim}
The theorem implies that the cutting polytope of $\bm{p}$ also contains $C(H)$ but 
excludes $\bm{p}$.
In other words, the intersection $\Pc(H) \cap \U(\bm{h})$ is 
a tighter relaxation of the convex hull of $C(H)$ compared with $\Pc(H)$.
Let $H'$ be the parity check matrix  obtained by stacking $H$ and $\bm{h}$, namely,
\begin{equation}
H' \equiv \left(
\begin{array}{c}
H \\
\bm{h} \\
\end{array}
\right).
\end{equation}
Note that $\Pc(H') = \Pc(H) \cap \U(\bm{h})$ has also geometrical uniformity because $\Pc(H')$ is 
a fundamental polytope. This means that the cutting polytope cuts not only $\bm{p}$ 
but also some non-codeword vertices of $\Pc(H)$ which are geometrically equivalent to $\bm{p}$.
The fractional distance of $H'$, $\df(H')$, thus can be larger than 
the fractional distance $\df(H)$ because the point $\bm{p} \in \Gamma(H)$ is excluded  from 
the new fundamental polytope $\Pc(H')$. Furthermore, the multiplicity (i.e., size of $\Gamma(H)$)
can be reduced as well by eliminating a fractional vertex with the minimum $\ell_1$-weight.
Figure \ref{fig:cutting} illustrates the idea of the cutting polytope.

\subsection{Cutting plane method: an example}

In this subsection, we will examine the idea of the cutting plane method described in
the previous subsection thorough a concrete example.
Let $H$ be a parity check matrix of $(7,4,3)$ Hamming code:
\begin{equation}
	H = 
	\left( 
	\begin{array}{ccccccc}
		1 & 0 & 1 & 1 & 1 & 0 & 0 \\
		1 & 1 & 0 & 1 & 0 & 1 & 0 \\
		0 & 1 & 1 & 1 & 0 & 0 & 1
	\end{array}
	\right).
\end{equation}
In this case, we have the index sets:
\begin{eqnarray*}
\Supp(\bm{h}_1) \hspace{-3mm}&=&\hspace{-3mm} \{1,3,4,5\}, \\
\Supp(\bm{h}_2) \hspace{-3mm}&=&\hspace{-3mm} \{1,2,4,6\}, \\
\Supp(\bm{h}_3) \hspace{-3mm}&=&\hspace{-3mm} \{2,3,4,7\}
\end{eqnarray*}
and
\begin{eqnarray*} 
X(\bm{h_1}) \hspace{-3mm}&=&\hspace{-3mm} \{ \{1\},\{3\}, \{4\},\{5\}, \\
	& & \{3,4,5\},\{1,4,5\},\{1,3,5\},\{1,3,4\}   \}, \\
X(\bm{h_2}) \hspace{-3mm}&=&\hspace{-3mm} \{ \{1\},\{2\}, \{4\},\{6\}, \\
	& & \{2,4,6\},\{1,4,6\},\{1,2,6\},\{1,2,4\}   \}, \\
X(\bm{h_3}) \hspace{-3mm}&=&\hspace{-3mm} \{ \{2\},\{3\}, \{4\},\{7\}, \\
	& & \{3,4,7\},\{2,4,7\},\{2,3,7\},\{2,3,4\}   \}.
\end{eqnarray*}
The fundamental polytope of $H$ is the set of points in $[0,1]^7$ satisfying 
\begin{equation}
\sum_{j \in S} f_j + \sum_{j \in ( \Supp(\bm{h_i}) \setminus S)} (1- f_j) \le 3
\end{equation}
for any $i \in \{1,2,3\}$ and any $S \in X(\bm{h}_i)$.
From some computations (details of computation are described in the next section), 
we can obtain  
the set of $\ell_1$-minimum weight vertices of $\Pc(H)$(i.e., $\Gamma(H)$): 
\begin{small}
\begin{equation*}
	\left( 0, \frac{2}{3}, \frac{2}{3}, \frac{2}{3}, 0, 0, 0 \right),
	\left( \frac{2}{3}, 0, \frac{2}{3}, \frac{2}{3}, 0, 0, 0 \right),
	\left( \frac{2}{3}, \frac{2}{3}, 0, \frac{2}{3}, 0, 0, 0 \right).
\end{equation*}
\end{small}
Therefore, in this case, $\df(H)$ is equal to $2$.

Assume that we choose $\bm{h} = (1,0,1,0,0,1,1)$ as a redundant row that is the sum of 
the second and third rows of $H$. Let $\bm{p} = (0,2/3,2/3,2/3,0,0,0) \in \Gamma(H)$.
It is easy to check that
\begin{equation}
p_3 = \frac{2}{3} > \sum_{l \in \Supp(\bm{h}) \setminus \{3\}} p_l = 0
\end{equation}
holds where $\Supp(\bm{h}) = \{1,3,6,7\}$. This means that $\U(\bm{h})$ is a cutting polytope 
of $\bm{p}$. By stacking $H$ and $\bm{h}$, we get a new parity check matrix $H'$ whose 
fundamental polytope does not contain $\bm{p}$ as its vertex.
In a similar manner, continuing the above process 
(appending redundant rows to $H$ for cutting the vertices in $\Gamma(H)$), 
we eventually obtain a parity check matrix $H^*$:
\begin{equation}
	H^* = 
	\left( 
	\begin{array}{ccccccc}
		1 & 0 & 1 & 1 & 1 & 0 & 0 \\
		1 & 1 & 0 & 1 & 0 & 1 & 0 \\
		0 & 1 & 1 & 1 & 0 & 0 & 1 \\ \hline
		1 & 0 & 1 & 0 & 0 & 1 & 1 \\
		1 & 1 & 0 & 0 & 1 & 0 & 1 \\
		0 & 0 & 0 & 1 & 1 & 1 & 1 \\
		0 & 1 & 1 & 0 & 1 & 1 & 0
	\end{array}
	\right).
\end{equation} 
The fractional distance of $H^*$ is equal to $3$ which is strictly larger than
the fractional distance  of $H$ ($\df(H)=2$). The details of a way to find an appropriate redundant rows
will be discussed in the subsequent sections.
It has been observed that the vectors in  $\Gamma(H^*)$ are {\em integral};
namely,  they are the minimum weight codewords of  Hamming code.

\subsection{A greedy algorithm for cutting plane method} \label{ssec:algorithm}
The previous example on Hamming code suggests that 
iterative use of the cutting plane method for a given parity check matrix may yield
a parity check matrix with redundant rows which is better than the original one 
in terms of the fractional distance. The following greedy algorithm, called {\em greedy cutting plane algorithm},
is naturally obtained from the above observation.

\vspace{0.4cm}
\fbox{
\begin{minipage}{8cm}
\begin{description}
\item[\underline{Greedy cutting plane algorithm}]
\item[Step 1 ] Evaluate $\Gamma(H)$.
\item[Step 2 ] Pick up $\bm{p} \in \Gamma(H)$.
\item[Step 3 ] Find a redundant row $\bm{h}$ of $H$ which gives a cutting polytope of $\bm{p}$.
	If such $\bm{h}$ does not exist, exit the procedure.
\item[Step 4 ] Update $H$ by
\begin{equation}
H := \left(
\begin{array}{c}
H \\
\bm{h} \\
\end{array}
\right).
\end{equation}
\item[Step 5 ] Return to Step 1.
\end{description}
\end{minipage}
}
\vspace{0.4cm}

Details on the process of finding a redundant row $\bm{h}$ is shown in the next section.
The most time consuming parts of the above algorithm are evaluation of $\Gamma(H)$ and
search for a redundant row. In the next section, we will discuss efficient implementations for 
these parts that are indispensable to deal with codes of long length.

\section{Efficient implementation}

\subsection{Efficient computation of $\df(H)$ and $\Gamma(H)$}

As described in the previous section, evaluation of $\df(H)$ and $\Gamma(H)$ is required for 
finding a redundant parity check matrix with better fractional distance.
An algorithm for computing $\df(H)$ has been proposed by Feldman \cite{Feldman}.

Firstly, we review the Feldman's method.
For any $k \in \{ 1, \ldots, M \}$,  Let $d_k$ be
\begin{equation}
	d_k \equiv \mini \sum_{i=1}^n f_i \st \bm{f} \in ( \Pc(H) \cap \Fc_k )
\end{equation}
where this LP problem is denoted by $LP_k$. Thus, $d_k$ can be considered as 
the $\ell_1$-weight of the minimum weight vertex on the facet $\Fc_k$ of $\Pc(H)$.
Since there exists at least one facet  of $\Pc(H)$
which includes $\bm{p}$ for  any vector $\bm{p}$ in $\Gamma(H)$, it is evident that
\begin{equation} \label{LPdfrac}
\df(H) = \min_{k = 1}^M \delta_k
\end{equation}
holds where $\delta_k = d_k$ if $d_k > 0$; otherwise $\delta_k = \infty$.
The LP problems $LP_k$ can be efficiently solved with an LP solver based on
the simplex algorithm or the interior point algorithm. From the solution of 
these LP problems, we can obtain the fractional distance of $H$ by (\ref{LPdfrac}).

The number of constraints related to a fundamental polytope defined by (\ref{equ:PcH})
 is an  exponential function of the row weight of $H$.
Thus the number of executions of the LP solver rapidly increases as
the row weight of $H$ grows.
Another formulation of the fundamental polytope proposed by Yang et al.\cite{Yang} can be used 
to reduce the number of constraints.
In their formulation, high weight rows of $H$ are divided to some low weight rows by introducing auxiliary variables.
Although their method is effective for evaluating the fractional distance as well, 
we here propose another efficient method for evaluating $\df(H)$ in this section.
In our method, a fundamental polytope is relaxed to a fundamental cone.
This method can be combined with Yang et al.'s formulation.

We need to prepare a relaxed version of the fractional distance before discussing 
another expression of the fractional distance.
For $k \in \Ffinact(H)$, let $d_k\rlx$ be
\begin{equation}
	d_k\rlx \equiv \mini \sum_{i=1}^n f_i \st \bm{f} \in ( \Kc(H) \cap \Fc_k ).
\end{equation}
This relaxed LP problem is denoted by $LP_k\rlx$. 
The relaxed fractional distance $d_k\rlx$ is defined by
\begin{equation}
	\df\rlx(H) \equiv \min_{k \in \Ffinact(H)} d_k\rlx . \label{equ:new_frac}
\end{equation}
The next theorem states a useful equivalence relation.
\begin{thim} \label{th:drac}
For a given $H$,  the following equality holds:
\begin{eqnarray}
	\df\rlx(H) = \df(H). \label{equ:thim1}
\end{eqnarray}
(Proof) See appendix.
\hfill $\square$
\end{thim}
A merit of Theorem \ref{th:drac} is that
the evaluation of $\df\rlx(H)$ 
takes less computational complexity than that of
the evaluation of $\df(H)$ using the Feldman's method. The reduction on 
the computational complexity comes from the following two reasons.

One reason is that  the feasible region of $LP_k\rlx$ is based on the fundamental cone $\Kc(H)$ (instead of  $\Pc(H)$)
which can be expressed with fewer linear constraints than  the fundamental polytope.
In the case of a regular LDPC code with row weight $w_r$, 
the number of linear constraints required to define $LP_k\rlx$ is $m w_r +1$.
On the other hand,  to express the fundamental polytope required in $LP_k$, 
$m 2^{w_r - 1}$  linear constraints are needed. 

Another reason of the reduction on complexity is that fewer executions of the LP solver are required for evaluating $LP\rlx_k$
because we can focus only on the inactive linear constraints in the case of $LP\rlx_k$.

\subsection{Search for redundant rows}

A straightforward way to obtain a redundant row $\bm{h}$ that gives a 
cutting polytope of a given point $\bm{p}$ is the exhaustive search.
Namely, each redundant row is checked 
whether it satisfies the condition (\ref{equ:h_add}) or not.
However, this naive approach is prohibitively slow even for codes of moderate length 
because there are $2^m$ redundant rows. We thus need a remedy 
to narrow the search space. The following theorem gives the basis of the 
reduction on the search space.
\begin{thim}\label{limited}
Assume that there exists a cutting polytope of $\bm{p} \in \Gamma(H)$.
The polytope $\U(\bm{h}^*)$ is such a cutting polytope of $\bm{p}$ and 
the redundant row $\bm{h}^*$ is given by
$
\bm{h}^* = \sum_{i=1}^m a_i^* \bm{h}_i.
$
Then, $\U(\bm{h})$ is also a cutting polytope of $\bm{p}$ where
$
\bm{h} = \sum_{i=1}^m a_i \bm{h}_i
$
and
\begin{equation}
a_i = 
\left\{
\begin{array}{cc}
a_i^* & i \in Q \\
0        & i \notin Q.
\end{array}
\right.
\end{equation}
The index set $Q$ is defined by
\begin{equation}
	Q \equiv \left\{ i \in \{ 1, \ldots, m \} : \exists j \in \Supp(\bm{p}), h_{ij} \ne 0 \right\}.
\end{equation}
(Proof) Let $j^*  \in \Supp(\bm{h}^*) \cap \Supp(\bm{p})$  be the index satisfying 
\begin{equation}
p_{j^*} > \sum_{l \in \Supp(\bm{h}^*) \setminus \{j^*\}}  p_l.
\end{equation}
Since $p_j = 0$ if $j \notin \Supp(\bm{p})$, the above condition is equivalent to 
\begin{equation}
p_{j^*} > \sum_{l \in (\Supp(\bm{h}^*) \cap \Supp(\bm{p}))  \setminus \{j^*\}}  p_l.
\end{equation}
From the definition of $\bm{h}$ and $Q$, we have
\begin{equation}
\Supp(\bm{h}^*) \cap \Supp(\bm{p}) = \Supp(\bm{h}) \cap \Supp(\bm{p}).
\end{equation}
This equality leads to the inequality
\begin{eqnarray} \nonumber
p_{j^*} &>& \sum_{l \in (\Supp(\bm{h}) \cap \Supp(\bm{p}))  \setminus \{j^*\}}  p_l \\
&=& \sum_{l \in \Supp(\bm{h})   \setminus \{j^*\}}  p_l.
\end{eqnarray}
The above inequality implies that $\U(\bm{h})$ is a cutting polytope of $\bm{p}$.
\hfill\qed
\end{thim}

The significance of the above theorem is that we can fix $a_i = 0$ for $i \notin Q$
in a search process without loss of the chance to find a redundant row
generating a cutting polytope.
Therefore, computational complexity to find a
redundant row can be reduced by using this property.
Let $V$ be a sub-matrix of $H$ composed from the columns of $H$ corresponding to
the support of $\bm{p}$. 
The index set $Q$ consists of the row indices of non-zero rows of $V$.
Thus, in the case of LDPC codes, 
the size of $Q$ is expected to be small when the size of $\Supp(\bm{p})$ is small
because of sparseness of the parity check matrices.
In such a case, the search space of the redundant rows are limited in a reasonable size.
For example, in the case of the LDPC code ``96.33.964'' \cite{Mackey}($n=96, m=48$, row weight 6, column weight 3),
the size of $\Supp(\bm{p})$ is $7 (\bm{p} \in \Gamma(H))$ and the size of $Q$ is $8$.

% change! %%%%%%%%%%%%%%%%%%%%%%%%%%%%%%%%%%%%%%%%%%%%%%%%%%%%%%%%%%%%%%%%%%%%%%%%
Let $H^Q$ be the $|Q| \times n$ sub-matrix of $H$ composed from a row vectors whose indices are included in $Q$. 
From Theorem 3, we can limit the search space to the linear combinations of rows of $H^Q$.
In the following, we will present an efficient search algorithm for a redundant row.
Let $\bm{\tau} = (\tau_1, \ldots, \tau_n)$ denote an indices vector that satisfies 
$p_{\tau_1} \ge p_{\tau_2} \ge \cdots \ge p_{\tau_n}$.

\vspace{0.4cm}
\fbox{
\begin{minipage}{8cm}
\begin{description}
\item[\underline{Redundant row search algorithm}]
\item[Step 1 ] Construct $H^Q$.
\item[Step 2 ] Permute columns of $H^Q$ to the following form: 
	\begin{equation*}
		H^Q \Pi = \left( 
		\begin{array}{ccccc}
			\bm{v}_{\tau_2} & \bm{v}_{\tau_3} & \cdots & \bm{v}_{\tau_n} & \bm{v}_{\tau_1}
		\end{array}
		\right),
	\end{equation*}
	where $\Pi$ denotes a column permutation matrix and $\bm{v}_j, j \in \{ 1, \ldots, n \}$ 
	denotes the $j$-th column vector of $H^Q$.
\item[Step 3 ] Convert $H^Q \Pi$ into $U$ of row echelon form by applying elementary row operations.
\item[Step 4 ] Let
	\begin{equation*}
		\hspace{-0.5cm}
		i^* \equiv \arg \min \{ i \in \{ 1, \ldots, |Q| \} : \bm{u}_i \mathrm{ \ satisfies \ } (\ref{equ:cnd}) \},
	\end{equation*}
	where $\bm{u}_i = (u_{i1}, \ldots, u_{in})$ denotes the $i$-th row vector of $U \Pi^{-1}$. 
\item[Step 5 ] Output $\bm{u}_{i^*}$.
\end{description}
\end{minipage}
}
\vspace{0.4cm}

The idea of the redundant row search algorithm is based on the fact that
$\bm{u}_i$ (a candidate of desirable redundant rows) tends to satisfy condition (\ref{equ:cnd}).
This can be explained as follows. Assume that $u_{i, \tau_1} = 1$.
From the definition of row echelon form,
\begin{equation*}
	u_{i, \tau_2} = u_{i, \tau_3} = \cdots = u_{i, \tau_K} = 0
\end{equation*}
holds where $K$ is an integer larger than or equal to $i-1$.
This means that
\begin{equation}
	|\Supp(\bm{u}_i) \cap \Supp(\bm{p})| \le |\Supp(\bm{p})| -i +1 \label{equ:bound_supp}
\end{equation}
holds for $i \in \{ 1, \ldots, |Q| \}$. Let $\eta_i$ be 
\begin{equation}
	\eta_i \equiv \sum_{l \in (\Supp(\bm{u}_i) \cap \Supp(\bm{p})) \setminus \{ \tau_1 \}} p_l.
\end{equation}
From the inequality (\ref{equ:bound_supp}), it is evident that $\{ \eta_1, \eta_2, \ldots \}$ is a 
decreasing sequence.
We thus can expect that condition (\ref{equ:cnd}), i.e., $p_{\tau_1} > \eta_i$, 
eventually holds as $i$ grows\footnote{Note that we cannot state that $\bm{u}_i$ satisfying (\ref{equ:cnd}) 
always exists because this argument is based on the assumption $u_{i, \tau_1} = 1$.}.
It may be reasonable to choose the smallest index $i$ satisfying (\ref{equ:cnd}) because such $\bm{u}_i$
would be sparser in the case of a low density matrix. A sparse redundant row is advantageous since it is 
able to cut other fractional vertices with small weight.
%%%%%%%%%%%%%%%%%%%%%%%%%%%%%%%%%%%%%%%%%%%%%%%%%%%%%%%%%%%%%%%%%%%%%%%%%%%%%%%%%%%

\section{Results}
\subsection{Application to Hamming, Golay and LDPC codes}
In this subsection, we applied the cutting plane method to Hamming code ($n=7, m=4$),
Golay code ($n=24, m=12$),
 and a short regular LDPC code ``204.33.484'' \cite{Mackey}
($n=204, m=102$, row weight 6, column weight 3).
We here use a parity check matrix of Golay code described in \cite{Schwartz}.
Let $\df$ be the fractional distance of the original parity check matrices
and  $\dfa$ be the fractional distance of parity check matrices generated by the cutting plane method.
Let $N_d$ be the number of the appended redundant rows.
The results are shown in Table \ref{tab:agsm_result}．
\begin{table}[htb]
	\centering
	\caption{Fractional distances of redundant parity check matrices obtained by the cutting plane method}
	\label{tab:agsm_result}
	\begin{tabular}{c|r|r|c|r|c|c} \hline \hline
		Code & $n$ & $m$ & $N_d$ & $\df$ & $\dfa$ & $d_{\min}$ \\ \hline
		Hamming & 7 & 3 & 4 & 2.000 & 3.000 & 3 \\
		Golay & 24 & 12 & 40 & 2.625 & 3.429 & 8 \\
		Golay & 24 & 12 & 100 & 2.625 & 3.895 & 8 \\
		LDPC & 204 & 102 & 9 & 5.526 & 5.646 & unknown \\ \hline
	\end{tabular}
\end{table}
For example, addition of 100 rows to the original parity check matrix of Golay code increases the
fractional distance from $2.625$ to $3.895$. It is expected that these matrices constructed 
by the cutting plane method shows better LP decoding performance compared with the original 
matrices because $\dfa$ is greater than $\df$ for all the cases.

\subsection{Decoding performances}
\begin{figure}[tbp]
	\centering
	\includegraphics[scale=0.6]{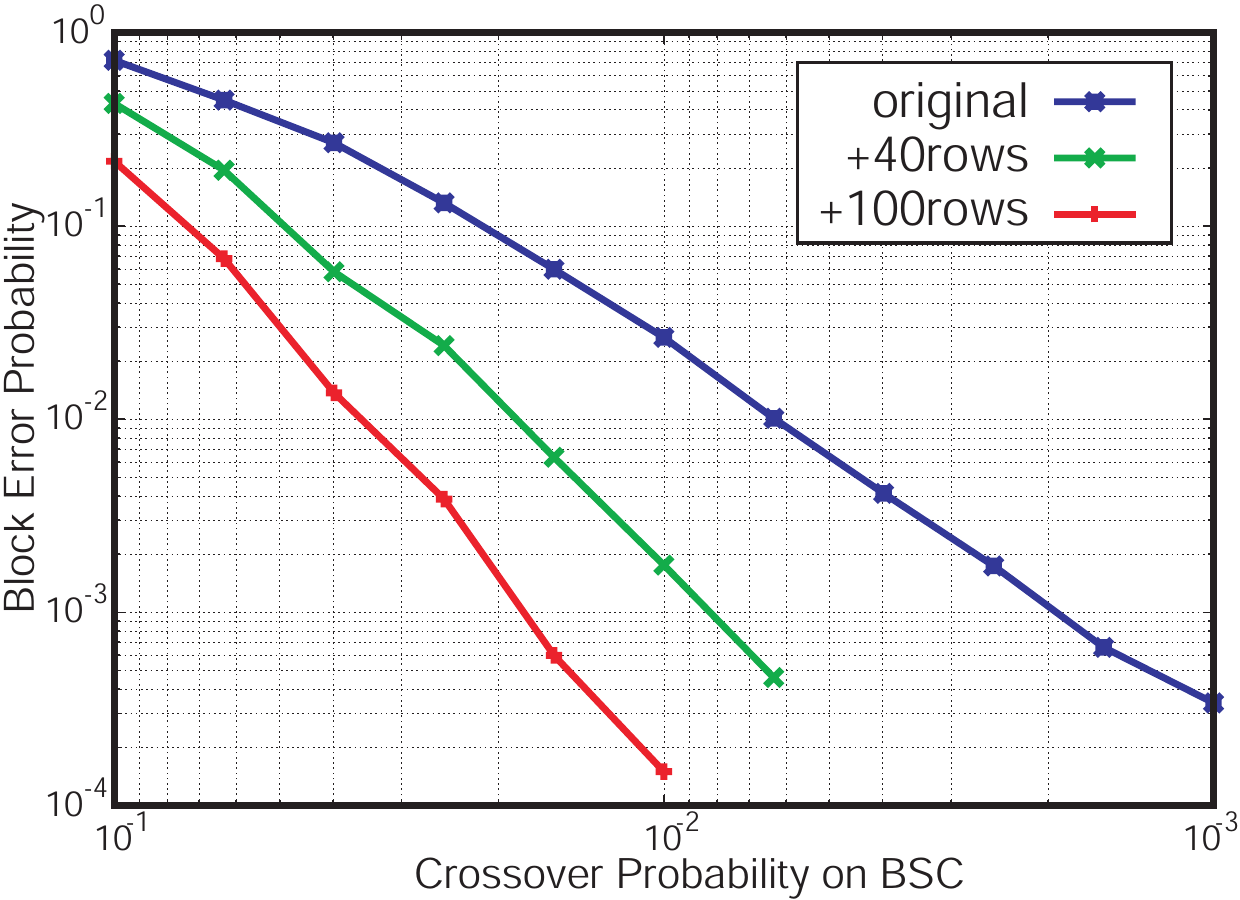}
	\caption{Comparison on block error probabilities of parity check matrices of Golay code (original $24 \times 12$ matrix, redundant $24 \times 52$ matrix, redundant $24 \times 112$ matrix )}
	\label{fig:golay_result_BSC}
\end{figure}

In this subsection, we will present decoding performance 
of redundant parity check matrices obtained by the cutting plane method.
We here assume BSC as a target channel and LP decoding \cite{Feldman} as a decoding 
algorithm used in a receiver.
Figure \ref{fig:golay_result_BSC} shows block error probabilities of 
the $24 \times 12$ parity check matrix of Golay code given in \cite{Schwartz} (labeled ``original"), 
the $24 \times 52$ parity check matrix with 40 redundant rows (labeled ``+40rows"),
and the $24 \times 112$ parity check matrix with 100 redundant rows (labeled ``+100rows'').

\begin{figure}[tbp]
	\centering
	\includegraphics[scale=0.6]{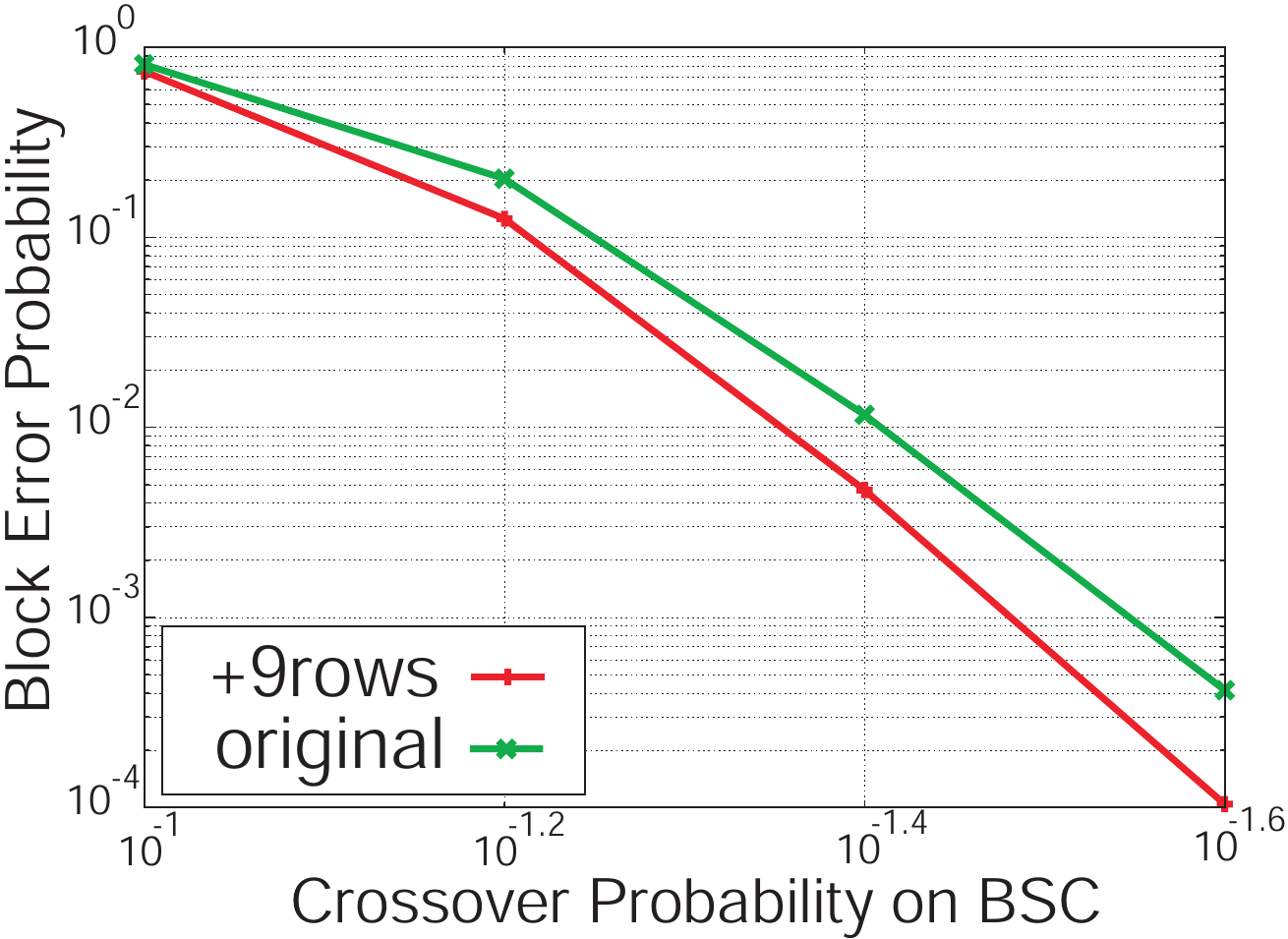}
	\caption{Comparison of block error probability of parity check matrices of the LDPC code ($n=204, m=102$)}
	\label{fig:204_result_BSC}
\end{figure}

From Figure \ref{fig:golay_result_BSC}, we can see that the block error probability of the
redundant matrix (+100rows) is 
approximately two order of magnitude lower than that of the original matrix when a crossover probability is $10^{-2}$.
Figure \ref{fig:204_result_BSC} shows block error probabilities of 
the original parity check matrix of ``204.33.484'' \cite{Mackey}
and that the parity check matrix with 9 redundant rows (labeled ``+9rows").
From Figure \ref{fig:204_result_BSC}, it can be observed that 
the slope of the error curve of  ``+9rows'' is steeper than 
that of ``original''.

\section{Conclusion}

In this paper, the cutting plane method based on redundant rows of a parity check matrix
for improving the fractional distance has been presented. In order to reduce the search complexity to find
an appropriate redundant row, we introduced an efficient technique to compute the fractional distance
and proved that the limited search space indicated in Theorem \ref{limited} is sufficient 
to find a redundant row generating a cutting polytope. 
Some numerical results obtained so far are encouraging.
The redundant parity check matrices constructed by the cutting plane method have larger 
fractional distance  than that of the original matrices. The simulation results support that
improvement on the fractional distance actually leads to better decoding performance under 
LP decoding.

\section*{Appendix}

\begin{prf}[Theorem 1]
For proving Theorem 1, we will prove the following two inequalities:
\begin{eqnarray} \label{ineq1}
\df\rlx(H) &\leq& \df(H), \\ \label{ineq2}
\df\rlx(H) &\geq& \df(H).
\end{eqnarray}

We first assume $\df\rlx(H) > \df(H)$ for proving inequality (\ref{ineq1})
by contradiction.
Assume that the index $i \in \{1,\ldots, M\}$ is given by
$
	i \in  \arg \min_{k=1}^M \delta_k.
$
Let $\bm{p}$ be the solution of $LP_i$. 
This means that $\bm{p} \in \Gamma(H)$ and 
$
	\bm{p} \in (\Pc(H) \cap \Fc_i) 
$
where $\Pc(H) \cap \Fc_i$ is the feasible set of $LP_i$.
Note that $\bm{p}$ is not the origin $0^n$ because $\delta_k = \infty$ holds when $\bm{p} = 0^n$.
Thus, $\Const_i$ is an inactive constraint, namely $i \in \Ffinact(H)$.
Since the fundamental corn $\Kc(H)$ contains the fundamental polytope $\Pc(H)$ as a subset,
the relation 
$
	\bm{p} \in (\Kc(H) \cap \Fc_i). 
$
holds as well. The intersection $\Kc(H) \cap \Fc_i$ is the feasible set of $LP_i\rlx$.
Let  the solution of $LP_i\rlx$ be  $\bm{p}^*$.
Since the objective function of $LP_i$ is identical to that of $LP_i\rlx$ and
the feasible set of $LP_i\rlx$ includes that of $LP_i$,
the $\ell_1$-weight of $\bm{p}^*$ is smaller than or equal to the $\ell_1$-weight of $\bm{p}$.
This implies $\df\rlx(H) \le \df(H)$ but it contradicts the assumption.
The proof of the inequality (\ref{ineq1}) is completed.

We next assume $\df\rlx(H) < \df(H)$ for proving inequality (\ref{ineq2}) by contradiction.
Assume that
$
	i \in  \arg \min_{k \in \Ffinact(H)} d_k\rlx .
$
Let $\bm{p}\rlx$ be the solution of $LP_i\rlx$.
From the definition of $LP_i\rlx$, it is evident that
$
\bm{p}\rlx \in  ( \Kc(H) \cap \Fc_i ) 
$
holds.
In the following, we will discuss the two cases: (i) $\bm{p}\rlx \in \Pc(H)$, (ii) $\bm{p}\rlx \notin \Pc(H)$.
We start from case (i). If $\bm{p}\rlx \in \Pc(H)$ holds, then  $\bm{p}\rlx \in  ( \Pc(H) \cap \Fc_i )$
holds because of the relation $\Pc(H) \subset \Kc(H)$.
Due to almost the same argument,  we obtain $\df\rlx(H) = \df(H)$ which contradicts the assumption.
We then move to case (ii). 
If $\bm{p}\rlx \notin \Pc(H)$ holds, there exists
$\Const_l$ satisfying
\begin{equation}
	\bm{p}\rlx \notin \Hc_l , \quad 0^n \in \Hc_l , \quad l \in \Ffinact(H). \label{equ:Cl}
\end{equation}
This is because  $\bm{p}\rlx \in \Kc(H)$ but $\bm{p}\rlx \notin \Pc(H)$.
Let $\Seg(\bm{p}\rlx)$ be the line segment between $\bm{p}\rlx$ and the origin:
\begin{equation}
	\Seg(\bm{p}\rlx) \equiv \left\{ \bm{p} \in R^n : 0 < t < 1,\  \bm{p} = t \bm{p}\rlx \right\}. \label{equ:line}
\end{equation}
The line segment $\Seg(\bm{p}\rlx)$ passes through $\Fc_l$.
Thus, there exists  the point $\bm{p}_l \in ( \Fc_l \cap \Seg(\bm{p}\rlx) )$.
Note that the line segment $\Seg(\bm{p}\rlx)$ is totally included in $\Kc(H)$ because $\bm{p}\rlx \in \Kc(H)$.
This leads to  $\bm{p}_l \in (\Kc(H) \cap \Fc_l)$ and it means that $\bm{p}_l$ is included in the feasible set of $LP_l\rlx$.
The $\ell_1$-weight of $\bm{p}_l$ is smaller than that of $\bm{p}\rlx$. This implies that 
the $\ell_1$-weight of  $\bm{p}_l$
is smaller than $\df\rlx(H)$.
However, it contradicts the definition of $\df\rlx(H)$.
The proof of the inequality (\ref{ineq2}) is completed.
\hfill \qed
\end{prf}

\section*{Acknowledgment}
We would like to thank the anonymous reviewers of Turbo Coding 2008 for helpful suggestions to 
improve the presentation of the paper.
This work was supported by the Ministry of Education, Science, Sports
and Culture, Japan, Grant-in-Aid for Scientific Research on Priority Areas
(Deepening and Expansion of Statistical Informatics) 180790091
and a research grant from SRC (Storage Research Consortium).

\end{document}